\documentclass[prl,twocolumn]{revtex4}
\usepackage{amsfonts}
\usepackage{pstricks}
\usepackage{pst-poly}
\usepackage{graphicx}
\usepackage{txfonts}

\newcommand{\inF}{\mbox{inF}}
\newcommand{\beq}{\begin{equation}}
\newcommand{\eeq}{\end{equation}}
\newcommand{\beqa}{\begin{eqnarray}}
\newcommand{\eeqa}{\end{eqnarray}}
\newcommand{\beqar}{\begin{eqnarray*}}
\newcommand{\eeqar}{\end{eqnarray*}}

\newcommand{\ie}{{\it i.e.,}\ }
\newcommand{\eg}{{\it e.g.,}\ }

\def\bi{\begin{itemize}}
\def\ei{\end{itemize}}
\def\be{\begin{equation}}
\def\ee{\end{equation}}
\def\bea{\begin{eqnarray}}
\def\eea{\end{eqnarray}}
\def\ben{\begin{eqnarray*}}
\def\een{\end{eqnarray*}}

\def\>{\rangle}
\def\<{\langle}

\newcommand{\mE}{{\mathcal E}}

\newcommand{\mX}{{\mathcal X}}
\newcommand{\mY}{{\mathcal Y}}
\newcommand{\mZ}{{\mathcal Z}}
\newcommand{\mG}{{\mathcal G}}

\newcommand{\mI}{{\mathcal I}}

\newcommand{\1} I 

\newcommand{\bra}[1]{\langle #1 |}
\newcommand{\ket}[1]{| #1 \rangle}

\def\*{\star}

\def\bar{\overline}

\begin{document}

\title{Quantum Error-Correcting Codes with Preexisting Protected Qubits}
\author{Ying Dong$^1$, Xiuhao Deng$^1$, Mingming Jiang$^1$, Qing
Chen$^1$, and Sixia Yu$^{1,2}$}\email{yusixia@ustc.edu.cn}
\affiliation{$^1$Hefei National Laboratory for Physical Sciences at
Microscale and Department of Modern Physics, University of Science
and Technology of China, Hefei 230026, P.R. China \\
$^2$Department of Physics, National University of Singapore, 2
Science Drive 3,  Singapore 117542}
\date{\today}

\begin{abstract}
We provide a systematic way of constructing entanglement-assisted
quantum error-correcting codes via graph states in the scenario of
preexisting perfectly protected qubits. It turns out that the
preexisting entanglement can help beat the quantum Hamming bound and
can enhance (not only behave as an assistance) the performance of
the quantum error correction. Furthermore we generalize the error
models to the case of not-so-perfectly-protected qubits and
introduce the quantity {\em infidelity} as a figure of merit and
show that our code outperforms also the ordinary quantum
error-correcting codes.
\end{abstract}


\maketitle

The quantum error-correcting code (QECC)
\cite{shor,7-q,5-q,5-q2,correct-condition} is an active way to deal
with the errors caused by the quantum noises during the process of
quantum communication and quantum computation. Simply speaking, a
QECC is just a subspace that corrects errors. The first quantum code
is the well known Shor's 9-qubit code \cite{shor}, which is a
quantum analog of the classical repetition code followed by Steane's
7-qubit code \cite{7-q} and the optimal 5-qubit code
\cite{5-q,5-q2}. Along with the establishment of stabilizer
formalism \cite{stab1,stab2,stab3} in QECC theory, various more
efficient quantum codes were constructed.

The constructions of QECCs depend on the error models. Standard
QECCs deal with the error model in which every qubit may equally go
wrong. Often there may exist some special qubits that are protected
from noises somehow, e.g., in a communication scenario in which
Alice want to send some qubits via a noisy quantum channel to Bob
while they may share some ideal EPR pairs beforehand. In that case
the qubits in Bob's hand are free from errors caused by the noisy
channel. For another instance, in the nuclear spin-electron spin
system, the error probability of nuclear spin is as $10^{-6}$ as
that of electron spin. Or some of our physical qubits are protected
by some QECCs already. It is therefore reasonable to assume that
some physical qubits in which our quantum data is encoded are
perfectly protected from errors. The entanglement-assisted QECC
(EAQECC) \cite{eaqecc} deals with exactly such a situation where the
perfectly protected qubits are ensured by the preexisting EPR pairs.
It is a special example of QECCs with preexisting protected qubits
dealing with an error model in which there are some physical qubits
that suffer errors with a smaller probability than other physical
qubits.

Recently a graphical approach to the construction of QECCs
\cite{gqecc} has been developed in the cases of of both stabilizer
and nonadditive codes \cite{rs,9123}, binary and nonbinary codes
\cite{nonbinary,looi}. For the binary case a codeword stabilized
codes approach \cite{zbei} has also been introduced. In this Letter
we shall at first present a graphical construction of all the
EAQECCs and then discuss in details two special codes found via our
approach: a family of codes beating the quantum Hamming bound and a
9-qubit code that demonstrates the fact that the entanglement can
really enhance the performance of the QECCs and not only behave as
an assistance. At last we consider the error model where the qubits
are not so perfectly protected and introduce the concept of {\it
infidelity} to characterize the performance of a code in such an
error model and discuss the advantage of our 9-qubit code beyond
EAQECC.

\paragraph{Graphical constructions of EAQECCs}
Instead of in a communication scenario, we shall develop our
graphical approach to the EAQECCs in the scenario where, among $n+e$
physical qubits, there exist $e$ {\it pure} qubits that suffer no
error at all during the whole quantum process. As we will note later
there is a slightly difference (one-way classical communications)
between these two scenarios.

Considering a graph $G=(V,\Gamma)$ composed of a vertex set $V$ with
$n+e$ vertices and edges specified by an {\it adjacency matrix}
$\Gamma$, which is a symmetric matrix with vanishing diagonal
entries and $\Gamma_{ab}=1$ if $a,b$ are connected and
$\Gamma_{ab}=0$ otherwise. Let $N_a=\{b\in V|\Gamma_{ab}=1\}$ denote
the neighborhood of vertex $a$ and $P$ be a subset of $V$ containing
$e$ vertices.  We label a system of $n+e$ qubits with $V$ and the
pure qubits with $P$. The graph state
\cite{graph-state1,graph-state2,graph-state3} on $G$ reads
\begin{equation}
|\Gamma\rangle=\prod_{a,b\in V}(\mathcal
U_{ab})^{\Gamma_{ab}}|+\rangle_x^V,
\end{equation}
where $\mathcal U_{ab}$ is the controlled phase gate between qubit
$a$ and $b$, i.e., $\mathcal U_{ab}=(1+\mathcal
Z_a+\mZ_b-\mZ_a\mZ_b)/2$ and $|+\rangle^V$ is the joint +1
eigenstate of all $\mX_a$ ($a\in V$) with $\mZ,\mX,\mY$ denoting
three Pauli matrices. The graph state $|\Gamma\rangle$ is also the
joint +1 eigenstate of $n+e$ vertex stabilizers
\begin{equation}
\mG_a=\mX_a\prod_{b\in N_a}\mZ_b,\quad (a\in V)\equiv\mX_a\mZ_{N_a}.
\end{equation}
Obviously $\mG_S=\prod_{a\in S}\mG_a$ stabilizes also the graph
state for arbitrary $S\subseteq V$. By specifying a collection of
$K$ different vertex subsets $\{C_i\}_{i=1}^K$ the graph state basis
$\{|\Gamma_{C_i}\rangle\equiv\mZ_{C_i}|\Gamma\rangle\}_{i=1}^K$
spans a $K$ dimensional subspace, where we have denoted
$\mZ_C=\prod_{a\in C}\mZ_a$.

Given a graph $G$ on the $n+e$ vertices and an integer $1\le d\le n$
we define a $(d,e)$-purity set as
\begin{equation}
\mathbb S_d=\left\{S\subseteq V\big|(S\cup N_S)\cap
P=\emptyset,|S\cup N_S|<d\right\}
\end{equation}
and a $(d,e)$-uncoverable set as
\begin{equation}
\mathbb D_d=2^V-\left\{\delta\bigtriangleup
N_\omega\big|(\delta\cup\omega)\cap
P=\emptyset,|\delta\cup\omega|<d\right\}.
\end{equation}
Here we have denoted by $A\bigtriangleup B=A\cup B-A\cap B$ the
symmetric difference of two subsets $A$ and $B$, by $|S|$ the number
of the elements in set $S$, and by $N_S=\bigtriangleup_{v\in S}N_v$
the neighborhood of a vertex subset $S$.

A {\em coding clique} $\mathbb C_d^K$ of a given graph $G$  with a
pure points set $P$ is a collection of $K$ vertex subsets that
satisfies:
\begin{itemize}
\item[i)] $\emptyset\in\mathbb C_d^K$
\item[ii)] $|S\cap C|$ is even for all $S\in \mathbb S_d$
and $C\in \mathbb C_d^K$
\item[iii)] $C\bigtriangleup C^\prime\in \mathbb D_d$ for
all $C,C^\prime\in \mathbb C_d^K$
\end{itemize}
We denote by $(G,K,d;e)$ as the subspace spanned by graph state
basis $\{|\Gamma_{C}\rangle|C\in\mathbb C_d^K\}$. If the coding
cliques form a group with respect to the symmetric difference, then
we call the coding clique as coding group and denote the
corresponding subspace as $[G,k,d;e]$ with $K=2^k$. As in
Ref.\cite{eaqecc} we denote by $[[n,k,d;e]]$ an EAQECC of length $n$
and distance $d$ with preexisting $e$ pure qubits. As usual we also
denote by $[[n,k,d]]$ a standard stabilizer code on $n$ qubits of
distance $d$. We have

{\bf Theorem} {\it The subspace $(G,K,d;e)$ is an EAQECC
$((n,K,d;e))$ and $[G,k,d;e]$ is an $[[n,k,d;e]]$ code.}

{\bf Proof.} It is enough to prove that for any error $\mE_d$ that
acts nontrivially on less than $d$ impure qubits we can get
$\bra{\Gamma_C}\mE_d\ket{\Gamma_{C^\prime}}=f(\mE_d)\delta_{CC^\prime}$
 \cite{correct-condition}\cite{5-q} for all $C,C^\prime\in \mathbb C_d^K$.
Without lose of generosity we assume that $\mE_d=\mathcal X_\omega
\mathcal Z_\delta$ for some pair of subsets $\delta,\omega$ with
$(\delta\cup\omega)\cap P=\emptyset$ and $|\delta\cup\omega|<d$,
which represents that there are $\mX$, $\mY$, and $\mZ$ errors on
the qubits in $\omega-\delta\cap\omega$, $\omega\cap \delta$, and
$\delta-\omega\cap\delta$ respectively. When acting on the graph
state the error $\mE_d\propto\mathcal G_\omega \mathcal Z_\Omega$
can be replaced by phase flip errors $\mZ_\Omega$ on
$\Omega:=\delta\bigtriangleup N_\omega$. If $\Omega$ is empty then
$\delta=N_\omega$ and the error is proportional to $\mathcal
G_\omega$. In this case we have
$\mG_\omega|\Gamma_C\rangle=|\Gamma_C\rangle$ for all $C\in \mathbb
C_d^k$ because $|\omega\cap C|$ is even which stems from the fact
that $|\omega\cup N_\omega|<d$, \ie $\omega\in \mathbb{S}_p$ and
Condition 1. Thus the error behaves like a constant operator on the
coding subspace and can be neglected. If $\Omega$ is not empty then
$\Omega\notin\mathbb D_d\cup \emptyset$ because it is covered by
$(\delta,\omega)$ and $|\delta\cup\omega|<d$. As a result
$\bra{\Gamma_C}\mE_d\ket{\Gamma_{C^\prime}}\propto\bra{\Gamma}
\mZ_{C\bigtriangleup C^\prime}\mZ_\Omega\ket\Gamma=0$ for all
$C,C^\prime\in \mathbb C_d^k$ because condition 2 ensures that
$C\bigtriangleup C^\prime\ne \Omega$. Now we have proved the first
part of the theorem. Further more, if we have a $k$ dimensional
coding group of a graph $G$ which is generated by $\langle
C_1,C_2,\ldots,C_k\rangle$, then we have a code $(G,2^k,d)$
according to the proof above. Since $k$ constraints $|S\cup
C_i|=$even for $i=1,2,\ldots,k$ have exactly $n-k$ independent
solutions $\langle S_1,S_2,\ldots,S_{n-k}\rangle$, the stabilizer of
the code is generated by $\langle\mG_{S_i}\rangle_{i=1}^{n-k}$.
\hfill Q.E.D.

Since the EAQECCs are a stabilizer code on $n+e$ qubits and all the
stabilizer code can be constructed in the graphical way
\cite{gqecc}, it follows that all the EAQECCs can also be found in
this graphical way.

According to this theorem, we can construct EAQECC systematically as
follows. First, input a graph $G=(V,\Gamma)$ on $n+e$ vertices.
Second, choose a distance $d$ and compute the $(d,e)$-purity set
$\mathbb S_d$ and $(d,e)$-uncoverable set $\mathbb D_d$. Third, find
all the K-clique $\mathbb C_d^K$\cite{clique}, and then for every
clique we obtain a $(G,K,d;e)$ code, \ie an $((n,K,d;e))$ code. And
if the coding clique form a group with respect to the symmetric
difference, we will have an stabilizer $[[n,k,d;e]]$ code.

In Table I we have listed the best EAQECCs we have found, giving the
distance $d$ as a function of the block size $n+e$ and number of
encoded qubits with $e=1$. The entries with an asterisk mark the
improvements over the best former EAQECC.
\begin{table}[tbph]    \begin{center}
{\small\begin{tabular}{@{\extracolsep{0.13cm}}| c || c | c | c | c |
c | c | c | c | c | c |}
     \hline$(n+e) \backslash k $&$1$& $2$& $3$& $4$& $5$& $6$& $7$& $8$& $9$& $10$\cr\hline
$3$&2& 1& 1&  &  &  &  &  & &  \cr\hline $4$& $3^*$ & 2& 1& 1&  &  &
& &&
 \cr\hline $5$& $4^*$& 2& 2& 1& 1&  &  &  & &  \cr\hline $6$& $5^*$& $3^*$& 2& 2 & 1 & 1& &
&  &   \cr\hline $7$& $6^*$ & $3^*$ & 2 & 2& 2& 1 & 1  &  &  &
\cr\hline $8$& $7^*$ & 3 & 3 & 2 & 2 & 2 & 1 & 1  & &   \cr\hline
$9$& $8^*$ & 3 & 3 & $3^*$ & 2 & 2 & 2 & 1  &  1&  \cr\hline $10$&
$9^*$ & 4 & 3 & 3 & $3^*$ & 2 & 2 & 2 & 1&1 \cr\hline
\end{tabular}
}\end{center}\caption{e=1} \end{table}

\paragraph{EAQECCs beating the quantum Hamming bound}

The simplest  EAQECC found via our graphical approach is the 1-error
correcting code $[[3,1,3;1]]$ which encodes 1 logical qubit by 3+1
physical qubits, including one pure qubit. In comparison, to encode
one logical qubit at least 4+1 or 3+2 or 5 physical qubits have to
be used in the known EAQECC and the standard code $[[5,1,3]]$.

We consider the star graph $S_4$ on 4 vertices as shown in Fig.1(a)
and denote the corresponding graph state on 4 qubits as
$|S_4\rangle$. Here we have supposed that the qubit labeled with 0
suffers no error at all.  Given $d=3$ and $e=1$ we obtain that
$(d,e)$-purity set is empty and the graph $S_4$ admits a coding
group with 2 elements $\{\emptyset, \{1,2,3\}\}$. The subspace
spanned by
\begin{equation}
\left\{|S_4\rangle, \mZ_1\mZ_2\mZ_3|S_4\rangle\right\}
\end{equation}
is the $[[3,1,3;1]]$ code.  Three generators of the stabilizer of
the code and the syndromes for all 9 single qubit errors are listed
in Table II.
\begin{table}[tbph]
$$
\begin{array}{cccc|c|cccccccc}
\hline\hline
0&1&2&3& &\mX_1,\mX_2,\mX_3&\mZ_1&\mZ_2&\mZ_3&\mY_1&\mY_2&\mY_3          \cr\hline
 \mX &\mZ&\mZ&\mZ& \mG_0      &-&+&+&+&-&-&-\cr
 \mI&\mX&\mX&\mI & \mG_1\mG_2 &+&-&-&+&-&-&+\cr
 \mI&\mX &\mI&\mX& \mG_1\mG_3 &+&-&+&-&-&+&-\cr
\hline\hline
\end{array}
$$
\caption{The stabilizer of the code $[[3,1,3;1]]$.}
\end{table}
We see that the code is not pure since all three bit flip errors $\mX_k$ ($k=1,2,3$)
give rise to the same syndrome. As a result the
quantum Hamming bound \cite{hamming bound}, which imposes $2^k(3n+1)\leq 2^{n+e}$
on 1-error-correcting EAQECCs of length $n+e$,  is violated by
our $[[3,1,3;1]]$ code where $n=3$ and $k=e=1$.

\begin{figure}
\includegraphics{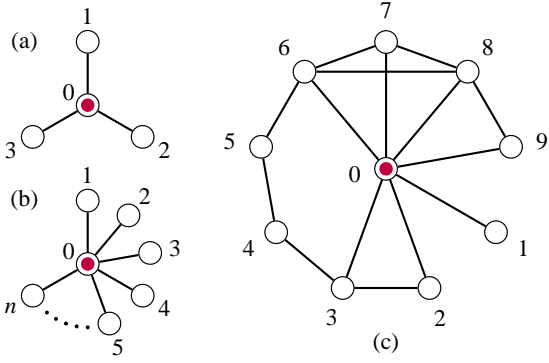} \caption{(a) The star graph for the code
$[[3,1,3;1]]$. (b) The star graph on $n+1$ vertices.  (c) The
coffeepot-like graph for the code $[[9, 5, 3; 1]]$. Here all the
purple vertices label the pure qubits. }
\end{figure}

In fact for any integer $n$ we are able to construct the code
$[[n,1,n;1]]$  with the star graph $S_{n+1}$ on $n+1$ qubits labeled
with numbers from 0 to n with the central qubit $0$ labeling the
pure qubit. This is the graphical code $[S_{n+1},1,n;1]$ with coding
group generated by $V-\{0\}$ and stabilizer generated by $\mG_0$ and
$\mG_1\mG_j$ with $j=2,3,\ldots,n$. It can be checked that all this
family of code except $n=4$ violate the quantum Hamming bound
$\sum_{s=0}^{t}3^sC_{n}^s\le 2^{n+e-k}$ with
$t=[\textstyle\frac{d-1}2]$ for the EAQECC $[[n,k,d;e]]$ since
$3^tC_{n}^t> 2^{n}$ for $n=d> 4$. It should be noticed that one-way
classical communications are needed to use the code $[[n,1,n;1]]$ in
the communication scenario. However with the presence of a one-way
classical communication channel the code $[[n,1,n;1]]$ is trivial
because Alice and Bob can transfer a qubit safely by teleportation
\cite{tele}.

\paragraph{An entanglement enhanced QECC} We consider the coffeepot-like
graph $T$ on 10 vertices as shown in Fig.1(c) and denote by
$|T\rangle$ the corresponding 10-qubit graph state. We suppose that
qubit 0 is perfectly protected from errors and label it with a
purple vertex. For $d=3$ and $e=1$ the graph $T$ admits a
5-dimensional coding group generated by
\begin{equation}
\big\{\{1,5,7\}, \{2,4,7\}, \{3,4,9\}, \{3,6,7\}, \{4,5,8\}\big\}.
\end{equation}
Accordingly the $2^5$-dimensional subspace spanned by the
graph-state basis ($\mu_1,\mu_2,\ldots,\mu_5=0,1$)
\begin{equation}
\mZ_1^{\mu_1}\mZ_2^{\mu_2}\mZ_3^{\mu_3+\mu_4}
\mZ_4^{\mu_2+\mu_3+\mu_5}\mZ_5^{\mu_1+\mu_5}\mZ_6^{\mu_4}\mZ_7^{\mu_2+\mu_4}
\mZ_8^{\mu_5}\mZ_9^{\mu_3}|T\rangle
\end{equation}
is a $[[9,5,3;1]]$ code whose stabilizer is generated by five graph
stabilizers as listed in Table III.
\begin{table}[tbph]
$$
\begin{array}{r|cccccccccc}\hline\hline
    n& 0&1& 2& 3& 4& 5& 6& 7& 8& 9\cr\hline
 \mG_0              &\mX& \mZ& \mZ& \mZ& \mI& \mI& \mZ& \mZ& \mZ& \mZ\cr
 \mG_1\mG_5\mG_8    &\mI& \mX& \mI& \mI& \mZ& \mX& \mI& \mZ& \mX& \mZ\cr
 \mG_2\mG_4\mG_8\mG_9 &\mZ& \mI& \mX& \mI& \mX& \mZ& \mZ& \mZ& \mY& \mY\cr
 \mG_3\mG_6\mG_9 &\mZ& \mI& \mZ& \mX& \mZ& \mZ& \mX& \mZ& \mI& \mX\cr
 \mG_1\mG_2\mG_6\mG_7&\mI& \mX& \mX& \mZ& \mI& \mZ& \mY& \mY& \mI& \mI\cr
\hline\hline
\end{array}
$$
\caption{The stabilizer of the code $[[9,5,3;1]]$.}
\end{table}
It is obvious that all single-qubit errors will lead to different
syndromes except three pairs of errors $\{\mZ_0,\mX_1\}$,
$\{\mX_0,\mZ_9\}$, and $\{\mY_0,\mY_6\}$. Therefore these could not
be corrected if qubit 0 were not perfectly protected from errors.

All the EAQECCs known so far either are identical to some QECCs, \eg
the code $[[7,3,3;1]]$ can be constructed from the stabilizer code
$[[8,3,3]]$, or are equivalent to protocols including standard QECCs
plus teleportation, \eg a $[[5,2,3;1]]$ code can be constructed via
the standard code  $[[5,1,3]]$ without using the protected qubit
together with encoding a logical qubit with the pure qubit, or in a
communication scenario, teleportating one qubit with the preexisting
ideal EPR pair. The $[[9,5,3;1]]$ code constructed above will
outperform both the standard QECC  and the QECC+teleportation, which
can encode at most 4 logical qubits.

We consider at first the communication scenario: Alice and Bob share
beforehand an ideal EPR pair and there is a noisy quantum channel
between them and Alice can send 9 qubits down the channel and it is
assumed that only 1 qubit in the channel will suffer errors which is
arbitrary and unknown.

By using an optimal stabilizer code $[[10,4,3]]$ \cite{stab3}, Alice
can encode 4 logical qubit in 10 qubits and send 9 qubits, keeping
her qubit in the ideal EPR pair, down the noisy channel to Bob.
After receive 9 qubits from Alice Bob can decode 4 logical qubits by
measuring a set of generators of the stabilize as in Table III on 10
qubits in his hand. As an alternative, Alice can also use an optimal
9-qubit stabilizer code $[[9,3,3]]$ to encode 3 logical qubits send
those 9 qubits down the noisy channel to Bob and then teleport one
qubit to Bob (one-way classical communications are needed). In both
protocols at most 4 qubits can be encoded.

On the other hand if Alice use the code $[[9,5,3;1]]$ instead she
can send 5 logical qubits to Bob. At first it obvious that Alice and
Bob can build the graph state $|T\rangle$ by local operations with
preexisting one ideal EPR pair. By local operations Alice can also
encode 5 logical qubits in 10 qubits in her hand and then send 9
qubits to Bob. It should be noticed that Alice can encode the
logical qubits without using her qubit in the EPR pair. In this way
Bob decode 5 qubits for 9 qubits Alice sent him and one qubit in EPR
pair. We see that in this EAQECC, the ideal EPR pair does not only
achieve its own task --- ensuring 1 qubit free from errors, but also
enhances the encoding ability of the other 9 qubits.

And then we consider the scenario of entanglement purification
\cite{purification} with one-way classical communications: Alice and
Bob share 9 copies of EPR among which 1 copy may go wrong but they
do not know which one and one ideal EPR pair. The best protocol
without the preexisting ideal EPR pair is to use the $[[9,3,3]]$
code (may also use $((9,12,3))$ \cite{9123} ) from which 3 ideal EPR
pairs can be purified. However if Alice and Bob measured the
stabilizer of the $[[9,5,3;1]]$ code instead on all 10 EPR pairs,
they can obtain 5 ideal EPR pairs. After extracting the preexisting
ideal EPR pair, they still have 4 ideal EPR pairs left. It is
equivalent to use a powerful $[[9,4,3]]$ code which does not exist.

\paragraph{Good-qubit-assisted QECCs}

In the discussions above, we have assumed an ideal error model in
which there preexist some perfectly protected qubits. It is however
more realistic to consider the error model in which there are some
physical qubits with a smaller error probability $p_e$ than the
error probability $p$ of other qubits. In this situation, we shall
demonstrate that the good-qubit-assisted QECC (GQAQECC) will
outperform the standard optimal codes as well. To do so we shall
introduce a reasonable figure of merit, namely the {\it infidelity},
to evaluate the performance of a code in addition to the distance
$d$ within the error model described above.

\begin{figure}
\begin{center}
\includegraphics[scale=0.27]{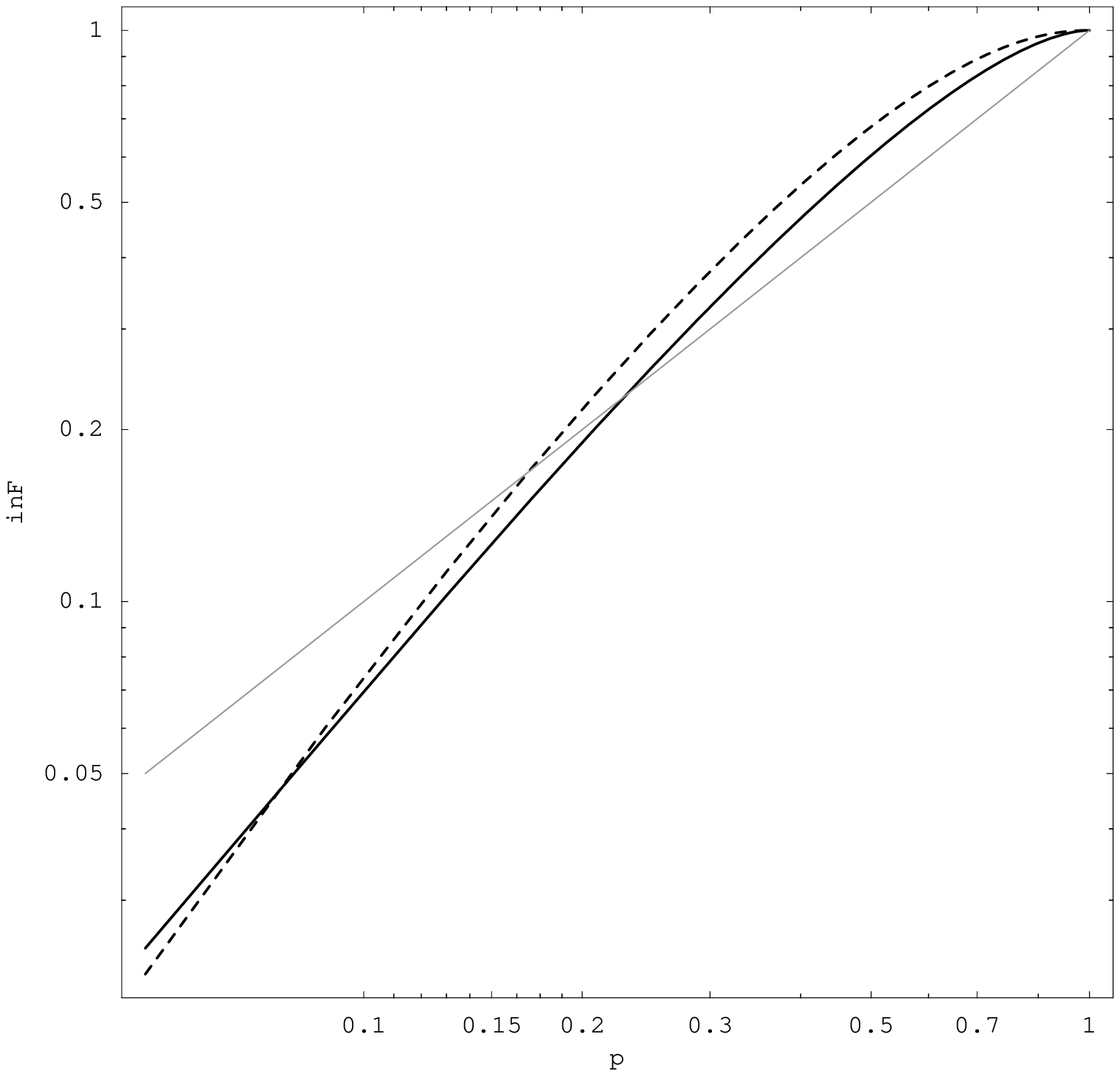}
\includegraphics[scale=0.27]{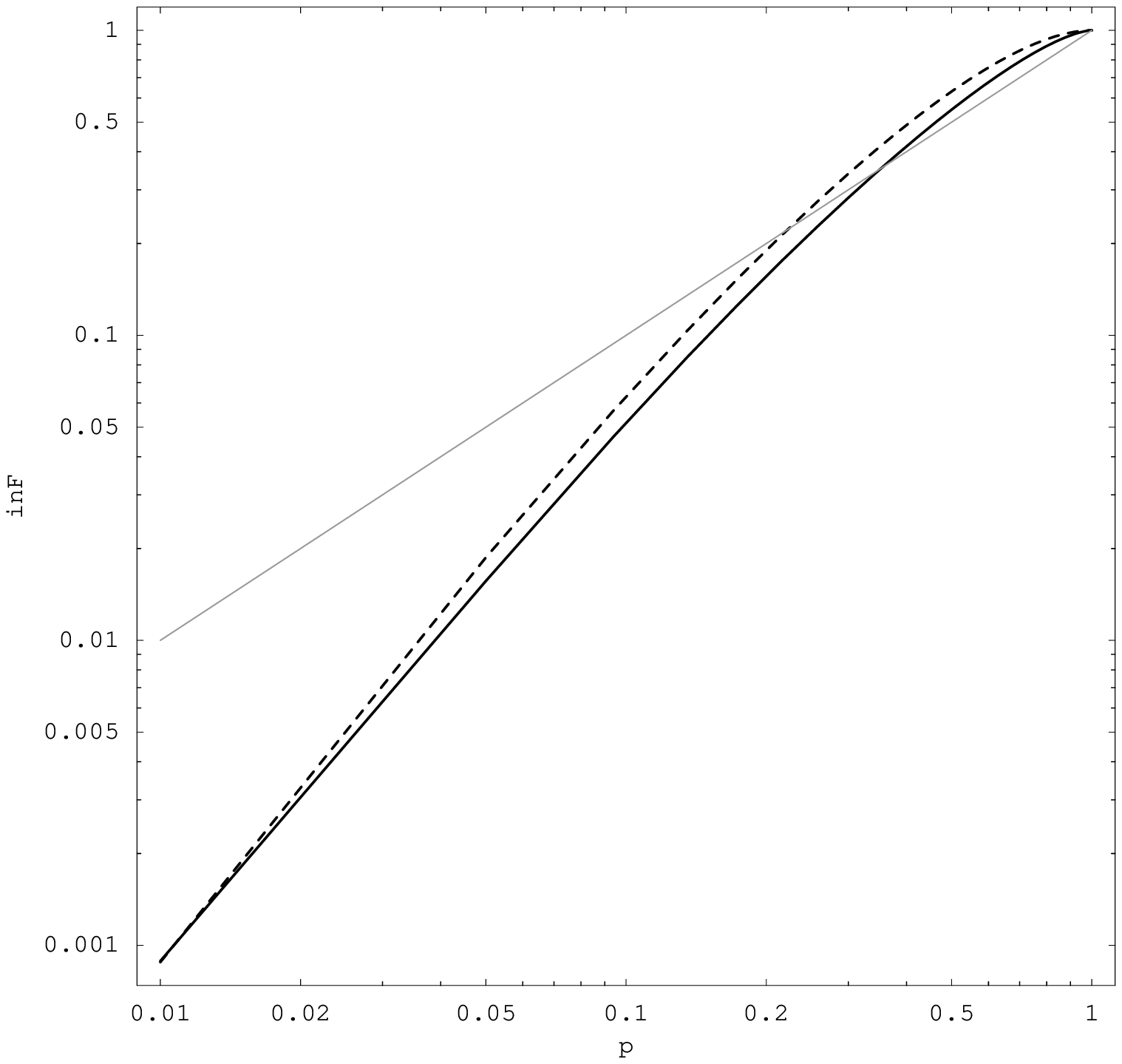} \caption{The infidelities of two codes
$[[10,4,3]]$ represented by the dashed curve and $[[9,5,3;1]]$
represented by the solid curve in the case of $p_e=p$ on lefthand
side and $p_e=p/10$ on the righthand side. The solid gray line
represents the infidelity in the case of no code is
used.}\end{center}
\end{figure}

A QECC  works perfectly only when correctable errors occur and we
denote by $P_C$ the probability of effective coding, i.e., the
probability of the occurrence of correctable errors. To compare
codes encoding different number of logical qubits, we image that we
used $k$ qubits directly instead of some code as logical qubits with
some error probability $p^\prime$ of physical qubits. This no-coding
scheme works perfect only if there is no error at all, i.e., the
probability of effective coding is $(1-p^\prime)^{k}$. To achieve
the same probability of effective coding as that of a QECC encoding
$k$ logical qubits, the error probability $p^\prime$ of the physical
qubits must be
\begin{equation}
\inF=1-(P_C)^{\frac1k}
\end{equation}
which is defined here to be the infidelity of a QECC. The smaller
the infidelity the better the code will perform. For examples
infidelities for the EAQECC $[[9,5,3;1]]$ and an optimal stabilizer
code $[[10,4,3]]$ read respectively
  \begin{eqnarray}
  \inF_{[[9,5,3;1]]}&=&1-\left(\bar p_e\bar p^9+9p\bar p_e\bar
  p^8\right)^{\frac15},\\
  \inF_{[[10,4,3]]}&=&1-\left(\bar p^9+9p\bar p_e\bar p^8\right)^{\frac14},
 \end{eqnarray}
where $\bar p=1-p$ and $\bar p_e=1-p_e$. In Fig.2 we have plotted
these two infidelities as functions of $p$ in the case of $p_e=p$
and $p_e=p/10$. We see that even in the symmetric case there are
some regions of $p$ that the entanglement enhanced code performs
better than the best standard QECC by encoding 1 more logical qubit
and with a less infidelity.

\paragraph{Conclusion}

With preexisting perfectly protected qubits we have demonstrated
that there are more efficient QECCs by constructing explicitly a
family of 1-error correcting codes violating the quantum Hamming
bound and a 9-qubit entanglement enhanced QECC that outperforms the
optimal standard QECC in both the communication scenario and
entanglement purification scenario. Within a more realistic error
model where there are qubits with smaller error probability than
other physical qubits, the GQAQECC performs also better than the
standard QECC basing one the infidelity as the figure of merit.

SXY acknowledges financial support from NNSF of China (Grant No.
10675107 and Grant No. 10705025), CAS, WBS (Project Account No):
R-144-000-189-305, Quantum information and Storage (QIS), and the
National Fundamental Research Program (Grant No. 2006CB921900).

\end{document}